%% file: main.tex
\documentclass[sigconf]{acmart} %UNCOMMENT FOR CAMERA READY
\usepackage{adjustbox, balance}
\usepackage{comment, url, multirow}
\AtBeginDocument{%
  \providecommand\BibTeX{{%
    Bib\TeX}}}

\copyrightyear{2025}
\acmYear{2025}
\setcopyright{cc}
\setcctype{by}
\acmConference[DFF '25]{Proceedings of the 1st Deepfake Forensics Workshop: Detection, Attribution, Recognition, and Adversarial Challenges in the Era of AI-Generated Media}{October 27--28, 2025}{Dublin, Ireland}
\acmBooktitle{Proceedings of the 1st Deepfake Forensics Workshop: Detection, Attribution, Recognition, and Adversarial Challenges in the Era of AI-Generated Media (DFF '25), October 27--28, 2025, Dublin, Ireland}\acmDOI{10.1145/3746265.3759668}
\acmISBN{979-8-4007-2047-5/2025/10}

\begin{document}

\title{Towards Reliable Audio Deepfake Attribution and Model Recognition: A Multi-Level Autoencoder-Based Framework}

\author{Andrea Di Pierno}
\orcid{0002-6279-9212}
\affiliation{%
  \institution{IMT School of Advanced Studies}
  \city{Lucca}
  \state{Tuscany}
  \country{Italy}
}
\email{andrea.dipierno@phd.unict.it}

\author{Luca Guarnera}
\affiliation{%
  \institution{University of Catania}
  \department{Department of Mathematics and Computer Science}
  \city{Catania}
  \state{Sicily}
  \country{Italy}}
\email{luca.guarnera@unict.it}

\author{Dario Allegra}
\affiliation{%
  \institution{University of Catania}
  \department{Department of Mathematics and Computer Science}
  \city{Catania}
  \state{Sicily}
  \country{Italy}}
\email{dario.allegra@unict.it}

\author{Sebastiano Battiato}
\affiliation{%
  \institution{University of Catania}
  \department{Department of Mathematics and Computer Science}
  \city{Catania}
  \state{Sicily}
  \country{Italy}}
\email{sebastiano.battiato@unict.it}

\renewcommand{\shortauthors}{Andrea Di Pierno, Luca Guarnera, Dario Allegra, and Sebastiano Battiato}

\begin{abstract}
The proliferation of audio deepfakes poses a growing threat to trust in digital communications. While detection methods have advanced, attributing audio deepfakes to their source models remains an underexplored yet crucial challenge. In this paper we introduce LAVA (Layered Architecture for Voice Attribution), a hierarchical framework for audio deepfake detection and model recognition that leverages attention-enhanced latent representations extracted by a convolutional autoencoder trained solely on fake audio. Two specialized classifiers operate on these features: \textit{Audio Deepfake Attribution} (ADA), which identifies the generation technology, and \textit{Audio Deepfake Model Recognition} (ADMR), which recognize the specific generative model instance. To improve robustness under open-set conditions, we incorporate confidence-based rejection thresholds. Experiments on ASVspoof2021, FakeOrReal, and CodecFake show strong performance: the ADA classifier achieves F1-scores over 95\% across all datasets, and the ADMR module reaches 96.31\% macro F1 across six classes. Additional tests on unseen attacks from ASVpoof2019 LA and error propagation analysis confirm LAVA’s robustness and reliability. The framework advances the field by introducing a supervised approach to deepfake attribution and model recognition under open-set conditions, validated on public benchmarks and accompanied by publicly released models and code. Models and code are available at \href{https://github.com/adipiz99/LAVA-framework}{https://github.com/adipiz99/LAVA-framework}.
\end{abstract}



\maketitle   
\input{sec/1_intro}

\input{sec/2_sota}
\input{sec/3_methods}
\input{sec/4_experiments}
\input{sec/5_results}

\input{sec/6_discussion}

\input{sec/7_conclusions}

\balance
\bibliographystyle{ACM-Reference-Format}
\bibliography{main}

\end{document}

%% file: sec/1_intro.tex
\section{Introduction}
\label{sec:intro}

The rise of synthetic media generation techniques, particularly those based on deep learning, has led to the widespread emergence of \textit{deepfakes}, manipulated audio, video, or images that convincingly imitate real individuals \cite{amerini2025challenges, Pontorno_2024}. Among these, audio deepfakes have attracted increasing attention due to their potential to impersonate voices in high-stakes contexts such as voice authentication, political communication, or disinformation campaigns. For instance, fraudsters once used AI-generated speech to impersonate a company executive and steal \$35 million from a bank\footnote{\url{https://www.forbes.com/sites/thomasbrewster/2021/10/14/huge-bank-fraud-uses-deep-fake-voice-tech-to-steal-millions/}, Last accessed: 21 June 2025}. In another case, a fake voice of President Biden was used in a robocall to mislead voters ahead of the New Hampshire primaries\footnote{\url{https://www.reuters.com/world/us/fake-biden-robo-call-tells-new-hampshire-voters-stay-home-2024-01-22/}, Last accessed: 20 June 2025}.\\
While audio synthesis offers valuable benefits in fields such as accessibility, entertainment, and human-computer interaction, it also introduces serious risks to security and public trust. In particular, the proliferation of audio deepfakes fosters a growing form of \textit{impostor bias}~\cite{CASU2024301795}, in which the authenticity of genuine audio is increasingly questioned. This erosion of trust impacts critical domains including journalism, legal evidence, and personal communication. While recent studies have begun to explore audio deepfake attribution~\cite{müller2022attackerattributionaudiodeepfakes, sourceTracing2024, stan2025tada, 10800741}, they often tackle the problem from alternative perspectives, such as attacker identification, pipeline inference, or unsupervised clustering, typically in closed-set conditions. Attribution plays a fundamental role in digital forensics, enabling investigators to infer the underlying technology or model family used to produce a fake. However, the task is particularly challenging due to the diversity and constant evolution of generation methods \cite{firc2025stopa}, and the subtle nature of the artifacts they leave behind.\\
Since binary detection of audio deepfakes (real vs. deepfake) has already been extensively studied in the literature \cite{yi2023survey}, in this paper we shift our focus toward the attribution of synthetic content. In particular, we propose a multi-level architecture for audio deepfake attribution, called LAVA (Layered Architecture for Voice Attribution). Specifically, LAVA is built upon a deep convolutional autoencoder trained exclusively on fake audio, which extracts compact latent representations reused across two task-specific classifiers:
\begin{itemize}
    \setlength\itemsep{0.5em}
    \item \textbf{Level 1: Audio Deepfake Attribution (ADA)}:  Given an audio deepfake sample $A_i$, the $i$-th input to be analyzed, the objective is to attribute it to its source manipulation technology by selecting among known generation methods such as \textit{ASVspoof2021}, \textit{FakeOrReal}, or \textit{CodecFake}. 
    \item \textbf{Level 2: Audio Deepfake Model Recognition (ADMR)}: identifies the specific generator model, the same task addressed by Guarnera et al.~\cite{guarnera2022ontheExploitation, guarnera2024mastering} in the deepfake image domain for model attribution. In our framework, this level is activated only when the first-level classifier (ADA) attributes the sample to the CodecFake dataset, as it is the only dataset that includes labeled generator classes. The input audio $A_i$ is then processed by a dedicated classifier that assigns it to one of the six known codec classes.
\end{itemize}
In detail, we distinguish between:
\begin{itemize}
    \item \textbf{Deepfake Attribution}: the task of assigning a fake audio sample to a known generation method or synthesis pipeline (e.g., a dataset or manipulation technology);
    \item \textbf{Deepfake Model Recognition}: the task of identifying the specific generator model, defined by its architecture and parameters, responsible for synthesizing the audio, among a known set of alternatives.
\end{itemize}
Both classifiers share the same encoder backbone and include an attention module to reweight salient features in the latent space. The system incorporates a confidence-based rejection threshold to abstain from uncertain classifications, thus improving robustness under open-set conditions. To rigorously evaluate our architecture, we perform experiments on three publicly available datasets, \textit{ASVspoof2021}~\cite{avspoof}, \textit{FakeOrReal}~\cite{for}, and \textit{CodecFake}~\cite{codecfake}. We measure classification performance using standard metrics, such as accuracy and F1-score, and conduct detailed ablation studies on the attention modules. We also introduce two complementary tests:
\begin{itemize}
    \item An \textbf{error propagation analysis}, which quantifies how misclassifications at the ADA Level affect downstream ADMR decisions;
    \item A \textbf{generalization test}, evaluating both classifiers on synthetic audio from ASVspoof2019 LA~\cite{wang2020asvspoof2019largescalepublic}, a dataset not seen during training but semantically close to ASVspoof2021.
\end{itemize}
Finally, we compare our method with recent state-of-the-art approaches in ADMR tasks, and show that LAVA achieves competitive or superior results across multiple settings.\\
Our main contributions are as follows:
\begin{itemize}
    \item LAVA, a modular multi-level architecture for audio deepfake attribution, leveraging an autoencoder trained solely on fake audio.
    \item A framework based on two levels: Audio Deepfake Attribution and Audio Deepfake Model Recognition.
    \item An attention mechanism, integrated into each classifier, proved to be effective through controlled ablation studies.
    \item A rejection strategy based on confidence thresholds, enabling the system to reject out-of-distribution inputs.
\end{itemize}
The remainder of the paper is organized as follows. Section~\ref{sec:sota} reviews related work on audio deepfake detection and attribution. Section~\ref{sec:method} presents our proposed architecture and training strategy. Section~\ref{sec:experimentalsetup} outlines datasets and experimental settings. Section~\ref{sec:results} reports empirical results, including ablations and robustness tests. Section~\ref{sec:discussion} provides an in-depth analysis of the results, examines comparisons with prior work, and highlights the strengths of the proposed architecture, particularly its hierarchical design, attention-based encoding, and robustness to open-set conditions, while also discussing potential limitations. Section~\ref{sec:conclusion} concludes the paper and outlines future directions.

%% file: sec/2_sota.tex
\section{Related works}
\label{sec:sota}

\subsection{Audio Deepfake Detection}
Recent years have seen a growing interest in detecting synthetic audio, driven by the increasing realism of text-to-speech (TTS) and voice conversion (VC) systems. A common strategy involves converting waveforms into time-frequency representations such as Mel-Frequency Cepstral Coefficients (MFCCs) or Constant-Q Cepstral Coefficients (CQCCs), which are then used as input to either convolutional~\cite{TODISCO2017516} or dense~\cite{jain2024} neural architectures for binary classification. End-to-end models such as RawNet~\cite{jung2019} and x-vector-based systems~\cite{desplanques2020ecapa} have demonstrated strong performance on raw inputs. However, these methods typically focus on binary classification (real vs. fake), and struggle to generalize across unknown synthesis methods~\cite{muller2023evaluating}. Recent studies~\cite{müller2024harderdifferentunderstandinggeneralization, yi2023survey} have also highlighted generalization as a major open challenge, especially in cross-dataset or open-set scenarios. To address this, some works have begun to explore self-supervised features~\cite{stan2025tada, phukan2024finder}, showing promising results for more robust detection and transferability.

\begin{figure*}[t!]
    \centering
    \includegraphics[width=\linewidth]{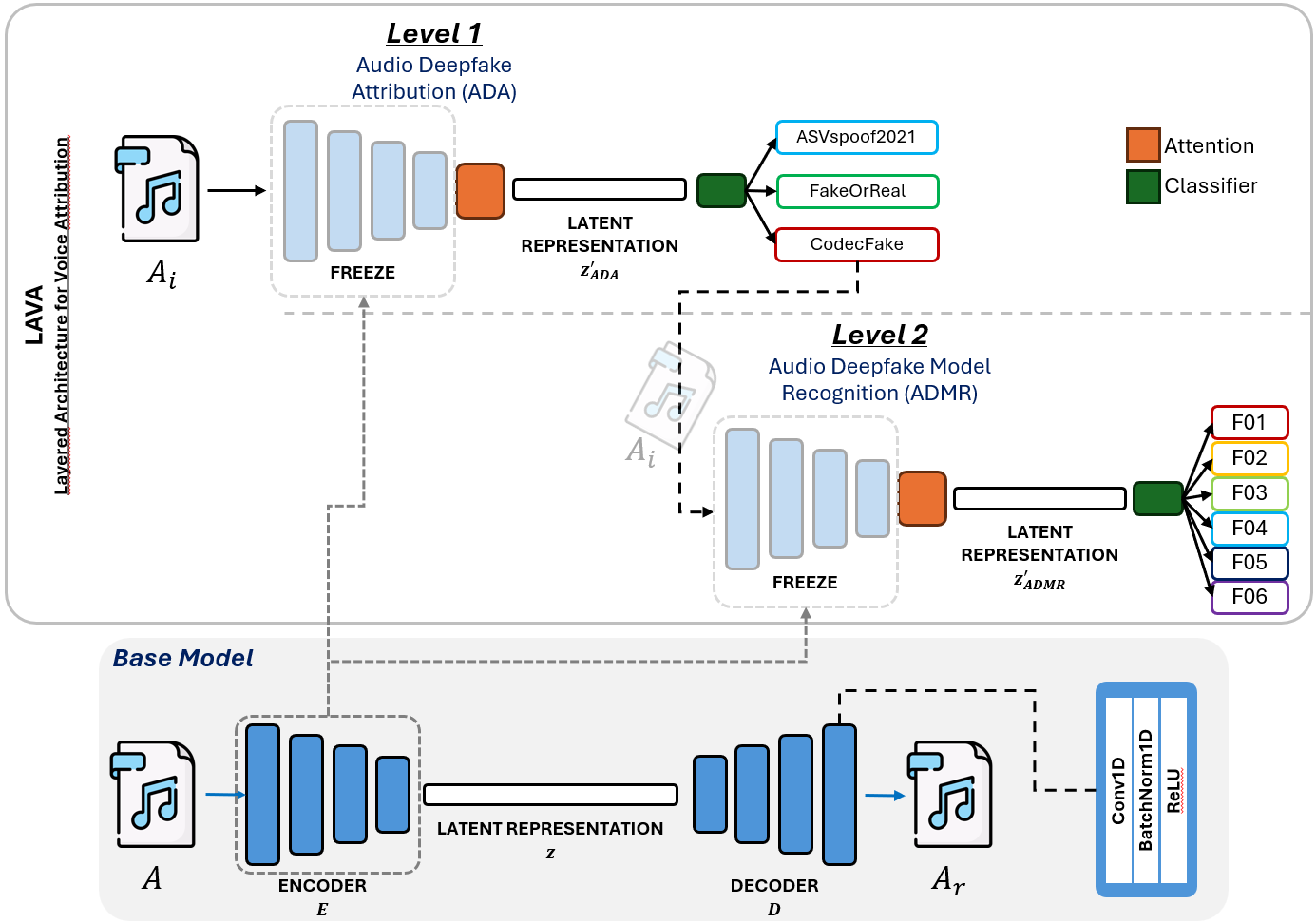}
    \caption{
    Overview of the LAVA framework. At the bottom, the \textit{base model} is a deep convolutional autoencoder trained to reconstruct fake audio inputs $A$ by minimizing the discrepancy between the original waveform $A$ and its reconstruction $A_r$, using a smoothed L1 loss. Once trained, the decoder is discarded and the encoder $E$ is reused as a frozen backbone for all subsequent classification tasks. At the top, an input audio sample $A_i$ is processed by the encoder to obtain a latent representation $z$, which is then passed through an attention module. In Level 1 (ADA), the resulting attended representation $z'_{\text{ADA}}$ is used to classify the sample into one of three dataset categories: \textit{ASVspoof2021 (ASV)}, \textit{FakeOrReal (FoR)}, or \textit{CodecFake (Codec)}. If the sample is attributed to \textit{CodecFake} and the classifier confidence exceeds a predefined rejection threshold, the sample is forwarded to Level 2 (ADMR). Here, the same encoder and attention module are reused to produce a second attended representation $z'_{\text{ADMR}}$, which is then classified into one of six codec-specific classes (\textit{F01–F06}). To build a robust attribution model, a threshold strategy was applied at each level of decision making: whenever the confidence associated with a prediction drops below a predefined threshold (different for each level), the corresponding sample is discarded and marked as ``unknown''.
    }
    \label{fig:architecture}
\end{figure*}

\subsection{Attribution in Multimedia Forensics}
Attribution is well-established in image forensics, where techniques identify source devices or editing tools~\cite{lakshmanan2019, cozzolino2018}. In audio, fewer works address generator attribution. CodecFake~\cite{codecfake} introduces a benchmark for codec-based manipulation detection. Klein et al. \cite{sourceTracing2024} explore the classification of spoofing system components through both end-to-end and two-stage learning strategies. Müller et al.~\cite{müller2022attackerattributionaudiodeepfakes} evaluated various models for audio deepfake attribution but observed rapid performance degradation in open-world setups. Yan et al.~\cite{yan2022audio} introduced the ADA dataset for audio deepfake attribution and proposed the CRML method, which enhances representation separation for open-set classification by leveraging multi-center learning. More recent proposals include the STOPA dataset~\cite{firc2025stopa}, designed to benchmark source tracing under systematic generation variation, and TADA~\cite{stan2025tada}, a training–free method leveraging SSL embeddings and k‑NN clustering for generator grouping. Wang et al.~\cite{wang2025adae} proposed attribution enhancement strategies to amplify synthesis-specific traces, and Negroni et al.~\cite{negroni2025source} reformulated attribution as a source verification problem using similarity learning. Neri et al.~\cite{ParalMGC2022} propose a dual-branch CNN that processes MFCC and GTCC features in parallel to attribute synthetic speech to its generation algorithm, achieving high accuracy on a closed-set dataset from the IEEE Signal Processing Cup.

\subsection{Our Contribution}
Differently from prior work, we propose a unified multi-level architecture tailored to deepfake audio attribution. It employs a shared autoencoder trained solely on fake samples to encode latent representations, followed by specialized classifiers for audio deepfake attribution (Level 1) and audio deepfake model recognition (Level 2). Attention mechanisms further refine these embeddings. To our knowledge, this is the first attempt to structure audio deepfake attribution as a hierarchical multi-task pipeline with built-in interpretability.

%% file: sec/3_methods.tex
\section{Proposed approach}
\label{sec:method}

In this section, we describe the multi-level architecture of our proposed framework for audio deepfake attribution and model recognition, and the datasets used for training and evaluation. We also discuss the rationale behind the ablation studies and our evaluation protocol. Figure~\ref{fig:architecture} illustrates an overview of the proposed approach.

\subsection{Datasets}
We use three publicly available datasets in our experiments:

\begin{itemize}
    \item \textbf{CodecFake}~\cite{codecfake}: A synthetic dataset composed of audio generated through six different speech codecs. It includes both real and fake utterances and is primarily used for fine-grained model-level attribution. The six fake classes differ in their compression strategies, architecture complexity, and training paradigms. For instance, \textit{SoundStream (F01)} \cite{soundstream} and \textit{EnCodec (F04)} \cite{encodec} are real-time neural codecs with transformer-based bottlenecks, while \textit{FuncCodec (F03)} \cite{funcodec} and \textit{AcademicCodec (F06)} \cite{academicodec} represent lightweight or academic baselines. \textit{SpeechToknizer (F02)} \cite{speechtokenizer} focuses on token-based speech modeling, and \textit{AudioDec (F05)} \cite{audiodec} employs diffusion-based reconstruction. These differences result in diverse signal characteristics and artifact patterns, making CodecFake suitable for evaluating model-level attribution capabilities.
    \item \textbf{ASVspoof2021}~\cite{avspoof}: A benchmark dataset for spoofing detection containing both bonafide and spoofed utterances, generated using a variety of synthesis techniques.
    \item \textbf{FakeOrReal (FoR)}~\cite{for}: A curated dataset designed for training and evaluating deepfake detection and attribution systems, containing real and synthetic audio segments.
\end{itemize}
All audio samples are converted to mono and resampled at 16~kHz to ensure uniformity across datasets. This sampling rate balances perceptual quality with computational efficiency and is commonly adopted in speech processing literature \cite{synder2018xvectors}. Waveforms are normalized by their peak absolute amplitude and trimmed or zero-padded to a fixed length of 3 seconds (i.e., 48,000 samples).\\ \\
%\subsubsection{Level 1: ADA}
As regards the first level (ADA) we use 75,000 fake samples evenly drawn from the three datasets, as shown in Table \ref{tab:ada_split}.
\begin{table}[ht]
\centering
\caption{Distribution of samples per dataset across training, validation, and test splits for ADA.}
\label{tab:ada_split}
\resizebox{\linewidth}{!}{%
\begin{tabular}{lcccc}
\toprule
\textbf{Split} & \textbf{CodecFake} & \textbf{ASVspoof2021} & \textbf{FakeOrReal} & \textbf{Total} \\
\midrule
Training   & 15,000 & 15,000 & 15,000 & \textbf{45,000} \\
Validation & 5,000  & 5,000  & 5,000 & \textbf{15,000} \\
Testing    & 5,000  & 5,000  & 5,000 & \textbf{15,000} \\
\midrule
\textbf{Total} & \textbf{25,000} & \textbf{25,000} & \textbf{25,000} & \textbf{75,000} \\
\bottomrule
\end{tabular}
}
\end{table}
\begin{comment}
\begin{itemize}
    \item \textbf{Training}: 45,000 samples (15,000 per dataset)
    \item \textbf{Validation}: 15,000 samples (5,000 per dataset)
    \item \textbf{Testing}: 15,000 samples (5,000 per dataset)
\end{itemize}
\end{comment}
\\
%\subsubsection{Level 2: ADMR}
The second level (ADMR) is trained on 313,282 fake samples from CodecFake, distributed as specified in Table \ref{tab:codec_split}.
\begin{table}[ht]
\centering
\caption{Class-wise distribution of CodecFake samples across training, validation, and test sets.}
\label{tab:codec_split}
\resizebox{\linewidth}{!}{%
\begin{tabular}{lccccccc}
\toprule
\textbf{Split} & \textbf{F01} & \textbf{F02} & \textbf{F03} & \textbf{F04} & \textbf{F05} & \textbf{F06} & \textbf{Total} \\
\midrule
Training   & 31,329 & 31,329 & 31,329 & 31,325 & 31,328 & 31,328 & \textbf{187,968} \\
Validation & 10,443 & 10,443 & 10,443 & 10,442 & 10,443 & 10,443 & \textbf{62,657} \\
Testing    & 10,443 & 10,443 & 10,443 & 10,442 & 10,443 & 10,443 & \textbf{62,657} \\
\bottomrule
\end{tabular}
}
\end{table}
\\
Unlike CodecFake, the ASVspoof2021 and FakeOrReal datasets do not include fine-grained labels specifying the exact generation model or codec used to synthesize each audio sample. They are organized as binary classification datasets with labels indicating only whether an utterance is real or fake. As a result, they are unsuitable for training or evaluating the ADMR classifier, which requires detailed ground truth annotations at the model level. For this reason, only CodecFake is used at Level 2 of the attribution pipeline.

\subsection{Proposed Autoencoder}

At the core of our architecture lies a convolutional autoencoder trained exclusively on fake audio samples. This design is based on the assumption that deepfakes, despite their variability, share generation-specific artifacts that can be encoded more effectively when real samples are excluded. Once training is complete, the decoder $D$ is discarded and only the encoder $E$ is kept and used in both ADA and ADMR modules. The encoder consists of a stack of convolutional layers, listed in Table~\ref{tab:encoder_architecture}, designed to progressively compress the input waveform into a latent representation $z$ of shape $(256, T')$, where $T'$ depends on the temporal downsampling rate. The encoder is trained by minimizing a Smoothed L1 Loss between the input $x$ and its reconstruction $\hat{x}$:

\begin{equation}
\mathcal{L}_{\text{smooth-L1}}(x, \hat{x}) = 
\begin{cases}
0.5 \cdot (x - \hat{x})^2 / \beta, & \text{if } |x - \hat{x}| < \beta \\
|x - \hat{x}| - 0.5 \cdot \beta, & \text{otherwise}
\end{cases}
\label{eq:smoothed_l1}
\end{equation}
where $\beta$ is a threshold hyperparameter. We use $\beta = 0.0001$ to emphasize small reconstruction errors while maintaining robustness to outliers.
\begin{table}[h]
\centering
\caption{Architecture of the encoder stack used to generate latent representations.}
\label{tab:encoder_architecture}
\resizebox{\columnwidth}{!}{%
\begin{tabular}{cccccc}
\toprule
\textbf{Layer} & \textbf{Type} & \textbf{In Channels} & \textbf{Out Channels} & \textbf{Kernel Size} & \textbf{Stride / Padding} \\
\midrule
1 & Conv1D     & 1   & 32  & 9 & 2 / 4 \\
2 & BatchNorm1D & -  & 32  & - & - \\
3 & ReLU       & -  & -   & - & - \\
4 & Conv1D     & 32  & 64  & 9 & 2 / 4 \\
5 & BatchNorm1D & -  & 64  & - & - \\
6 & ReLU       & -  & -   & - & - \\
7 & Conv1D     & 64  & 128 & 9 & 2 / 4 \\
8 & BatchNorm1D & -  & 128 & - & - \\
9 & ReLU       & -  & -   & - & - \\
10 & Conv1D    & 128 & 256 & 9 & 2 / 4 \\
11 & BatchNorm1D & - & 256 & - & - \\
12 & ReLU      & -  & -   & - & - \\
\bottomrule
\end{tabular}
}
\end{table}
\\
After training, all encoder weights are frozen except for the final convolutional layer, which remains trainable to support task-specific adaptation. The encoder thus acts as a shared feature extractor for both levels of the LAVA framework.\\
To enhance feature saliency, an attention mechanism is applied to the latent representation $z$, defined as:
\begin{equation}
z' = z \odot \sigma(\text{Conv1D}(z))
\label{eq:conv1d}
\end{equation}
where $\sigma$ is the sigmoid activation, $\odot$ denotes element-wise multiplication and $z'$ (called $z'_{\text{ADA}}$ for Level 1 and $z'_{\text{ADMR}}$ for Level 2) represents the reweighted latent representation obtained by modulating each channel of $z$ according to its learned relevance through the attention mechanism, as shown in Table \ref{tab:attention_classifier}.
The latent representation $z'_{\text{ADA}}$ or $z'_{\text{ADMR}}$ is then passed through a classifier composed of adaptive average pooling, two fully connected layers with ReLU activation, and a final linear layer that outputs logits for either 3 (ADA) or 6 (ADMR) classes.
\begin{table}[h]
\centering
\caption{Architecture of attention mechanism and classification head.}
\label{tab:attention_classifier}
\resizebox{\columnwidth}{!}{%
\begin{tabular}{lcc}
\toprule
\textbf{Component} & \textbf{Layer} & \textbf{Details} \\
\midrule
\multirow{2}{*}{\textbf{Attention}} 
  & Conv1D & In: 256, Out: 256, Kernel Size: 1 \\
  & Sigmoid & Element-wise activation over channels \\
\midrule
\multirow{4}{*}{\textbf{Classifier}} 
  & AdaptiveAvgPool1D & Output shape: (256, 1) \\
  & Flatten & Output shape: (256) \\
  & Linear + ReLU & In: 256, Out: 128 \\
  & Linear (Output) & In: 128, Out: 3 (ADA) or 6 (ADMR) \\
\bottomrule
\end{tabular}%
}
\end{table}
\\
This modular design enables encoder reuse while maintaining classifier independence. Additionally, the shared latent space fosters consistency across attribution levels and enables clearer task separation, which facilitates analysis and debugging of individual components.

\subsection{Level 1 - ADA}
The aim of the first level (ADA) is to determine the synthesis technology used to generate a given fake audio sample $A_i$. Specifically, the frozen encoder $E$ (pretrained as part of the base autoencoder and fine-tuned only in its final convolutional layer) processes $A_i$ to produce a latent representation $z$. This representation is refined through an attention mechanism to yield $z'_{\text{ADA}}$. The latent representation $z'_{\text{ADA}}$ is then classified into one of three dataset categories: \textit{ASVspoof2021 (ASV)}, \textit{FakeOrReal (FoR)}, or \textit{CodecFake (Codec)}. This stage serves as the entry point to the LAVA pipeline. A confidence-based rejection mechanism (Section~\ref{sec:rejection}) is applied to the softmax output: if the maximum confidence score is below the threshold $\tau_{\text{ADA}}$, the sample is rejected and labeled as unknown.

\subsection{Level 2 - ADMR}
In the second stage (ADMR) attribution proceeds only if the output of the ADA classifier corresponds to the \textit{CodecFake} class and the associated confidence score exceeds the threshold $\tau_{\text{ADA}}$. The same encoder $E$ and attention mechanism are reused to extract an attended latent representation $z'_{\text{ADMR}}$ from the original input $A_i$. This refined embedding is then processed by a second classifier to attribute the sample to one of six codec-specific generation classes (\textit{F01–F06}). A second rejection threshold $\tau_{\text{ADMR}}$ is applied at this level: if the classifier's confidence is below this threshold, the sample is rejected and labeled as ``unknown''. This mechanism serves to limit the propagation of erroneous predictions from Level 1 and prevent misclassification of samples that, although routed to ADMR, deviate from known latent patterns. Together, the modularity of $E$, the attention refinement, and the hierarchical rejection thresholds contribute to the pipeline’s robustness in open-set attribution scenarios.

\subsection{Ablation Settings}
To evaluate the impact of the attention mechanism, we train and test each classifier both with and without attention. The same training setup is used for all configurations: 50 epochs, batch size of 16, and early stopping based on validation loss.

\subsection{Rejection Threshold}
\label{sec:rejection}

To improve the system's robustness and its generalization capabilities, we adopt a rejection mechanism based on confidence scores. For each classifier, we compute a rejection threshold during training, defined as the minimum confidence score required for a prediction to be accepted. Specifically, this threshold is not arbitrarily fixed but is derived from the distribution of softmax confidence scores observed on the training set. For each training sample, we record the softmax confidence associated with its predicted class. These values are then sorted in descending order, and the threshold is set at the percentile that ensures at least 85\% classification accuracy on the training data. This ensures that only predictions made with sufficient confidence are accepted. At test time, predictions with a confidence score above this threshold are accepted as valid class predictions. Conversely, if the predicted class confidence falls below the threshold, the sample is rejected and assigned to an ``unknown'' class. This implies that the input is considered inconsistent with any of the known classes seen during training. The rejection mechanism is applied independently in both the ADA and ADMR classifiers and plays a crucial role in enabling open-set attribution and limiting error propagation across stages in the hierarchical architecture.

%% file: sec/4_experiments.tex
\section{Experimental Setup}
\label{sec:experimentalsetup}

In this section, we describe the training protocol, evaluation metrics, and implementation details used in our experiments. We also outline the baseline setup for the ablation studies introduced in Section~\ref{sec:method}.

\subsection{Autoencoder Pretraining}
\begin{figure}[t!]
    \centering
    \includegraphics[width=\linewidth]{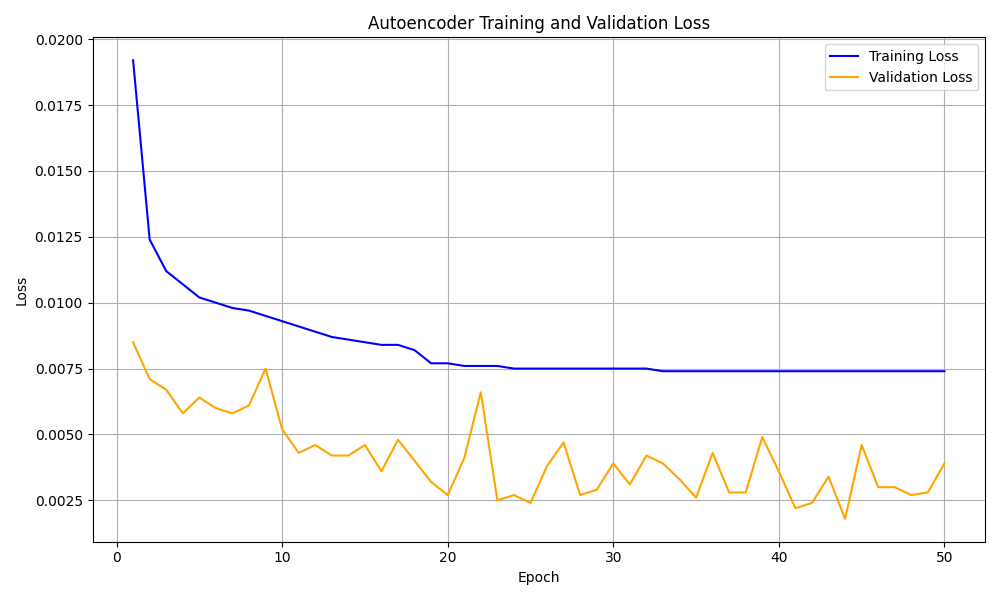}
    \caption{Training and Evaluation Loss of the autoencoder}
    \label{fig:autoencoder_curves}
\end{figure}

The shared encoder $E$, used in all classifiers, is derived from a deep convolutional autoencoder trained exclusively on fake audio samples from the CodecFake dataset. The dataset comprises 313{,}282 samples, evenly distributed across six codec classes. We employed a training/validation split of 80/20, resulting in 250{,}625 samples for training and 62{,}657 for validation. While the pretraining samples are drawn from the same dataset later used in the ADMR task, the autoencoder is trained solely for reconstruction without using generator labels. This ensures that representation learning remains disentangled from the downstream classification objectives. The autoencoder was trained for a maximum of 50 epochs using the Adam optimizer with a learning rate of $1 \times 10^{-4}$, weight decay of $1 \times 10^{-5}$, and batch size of 16. Early stopping was based on validation loss. The reconstruction objective is a Smoothed L1 Loss~(Eq.~\ref{eq:smoothed_l1}) with $\beta=0.0001$, which balances robustness to outliers with sensitivity to small deviations. The best model was selected at epoch 44, achieving a training loss of 0.0074 and a validation loss of 0.0018 (Figure \ref{fig:autoencoder_curves}). These values confirm that the encoder learned a compact and high-fidelity representation of fake audio samples, capturing generation-specific artifacts while filtering out irrelevant variance. The decoder $D$ is discarded after training, and the encoder is used as a frozen backbone in all downstream classifiers (with the exception of its final convolutional layer, which remains trainable).

\subsection{Training Procedure}
All classifiers are trained independently using the Adam optimizer with a learning rate of $1 \times 10^{-4}$ and weight decay of $1 \times 10^{-5}$. We adopt a batch size of 16 and a maximum of 50 training epochs with early stopping based on validation loss to prevent overfitting. The encoder weights are frozen during training, except for the final convolutional layer, which remains trainable to allow mild task-specific adaptation. All experiments are performed on a workstation equipped with an NVIDIA RTX A6000 GPU (48GB VRAM) and a 32-core AMD Ryzen Threadripper PRO 3975WX CPU. Datasets are preprocessed and stored as normalized waveforms of fixed length (3 seconds, 16~kHz, mono), as detailed in Section~\ref{sec:method}.

\subsection{Evaluation Protocol}
Each classifier is evaluated on the dedicated test split described in Section~\ref{sec:method}. We report standard classification metrics including Accuracy, Precision, Recall, and F1-score, both per class and as macro averages in Section \ref{sec:results}.\\
To assess the reliability and robustness of our attribution framework, we report not only standard classification results but also two additional evaluations: an error propagation test and a generalization test. The former simulates the full inference pipeline to quantify how misclassifications in the initial attribution stage (ADA) affect downstream model recognition (ADMR). The latter evaluates the system's capacity to handle unseen data. These analysis are reported in Section \ref{sec:results}.

\subsection{Ablation Protocol}
To isolate the contribution of the attention mechanism, we train and evaluate versions of each classifier with the attention module removed. All other components and hyperparameters remain unchanged. Performance is compared against the full-attention variants to quantify the impact of feature reweighting on attribution accuracy.

\begin{comment}
\subsection{Implementation Details}
We initialize the encoder weights from a convolutional autoencoder trained solely on fake audio samples from CodecFake. The decoder is discarded post-training. During classifier training, we freeze all encoder layers except the final convolution, which allows minor adjustments to the learned latent space. Each classifier includes a 1D attention block, adaptive average pooling, and two fully connected layers, as described in Section~\ref{sec:method}.\\
All training, validation, and test splits are balanced across classes. The ADA classifier is trained on 75,000 fake samples (25,000 from each dataset), while the ADMR classifier uses 313,282 fake samples from CodecFake, evenly distributed across six codec classes.
\end{comment}

%% file: sec/5_results.tex
\section{Results and Discussion}
\label{sec:results}

\subsection{Evaluation Metrics}
We evaluate our models using standard classification metrics: accuracy, precision, recall, and F1-score. These metrics provide a comprehensive overview of classification performance across both balanced and imbalanced class distributions.

\subsection{Experimental Results}
As shown in Tables~\ref{tab:ADA_with_attention} and~\ref{tab:ADMR_with_attention}, our models achieve strong performance across all attribution levels when using the attention mechanism. The ADA module achieves an overall accuracy of 96.21\%. Most notably, the ADMR model reaches a macro-average F1-score of 96.31\% across six codec classes, demonstrating the architecture’s effectiveness for fine-grained attribution. These results confirm that the integration of attention modules, as demonstrated in Section \ref{sec:ablation}, helps the model focus on salient latent features, improving class separability in the encoded space.

\begin{table}[b!]
\centering
\caption{Audio Deepfake Attribution (ADA) results}
\label{tab:ADA_with_attention}
\resizebox{\columnwidth}{!}{%
\begin{tabular}{lccc}
\toprule
\textbf{Dataset} & \textbf{Precision} & \textbf{Recall} & \textbf{F1-Score} \\
\midrule
CodecFake & 0.9749 & 0.9568 & 0.9658 \\
ASVspoof2021 & 0.9402 & 0.9720 & 0.9558 \\
FakeOrReal & 0.9724 & 0.9576 & 0.9649 \\
\midrule
Accuracy & \multicolumn{3}{c}{0.9621} \\
Macro Avg & 0.9625 & 0.9621 & 0.9622 \\
Weighted Avg & 0.9625 & 0.9621 & 0.9622 \\
\bottomrule
\end{tabular}%
}
\end{table}

\begin{table}[t!]
\centering
\caption{Audio Deepfake Model Recognition (ADMR) results}
\label{tab:ADMR_with_attention}
\resizebox{\columnwidth}{!}{%
\begin{tabular}{lccc}
\toprule
\textbf{Class} & \textbf{Precision} & \textbf{Recall} & \textbf{F1-Score} \\
\midrule
F01 & 0.9975 & 0.9980 & 0.9978 \\
F02 & 0.9016 & 0.9325 & 0.9168 \\
F03 & 0.9921 & 0.9959 & 0.9940 \\
F04 & 0.9789 & 0.9880 & 0.9835 \\
F05 & 0.9778 & 0.9741 & 0.9760 \\
F06 & 0.9319 & 0.8907 & 0.9108 \\
\midrule
Accuracy & \multicolumn{3}{c}{0.9632} \\
Macro Avg & 0.9633 & 0.9632 & 0.9631 \\
Weighted Avg & 0.9633 & 0.9632 & 0.9631 \\
\bottomrule
\end{tabular}%
}
\end{table}

\subsection{Error Propagation Test}
We simulate the full inference pipeline by feeding each sample through the ADA classifier and forwarding it to the ADMR classifier only when the ADA prediction corresponds to \textit{CodecFake}. This setup reflects real-world deployment, where upstream errors affect downstream attribution. The evaluation was conducted on 21,000 samples, including 15,000 fake (5,000 per dataset's test set) and 6,000 randomly selected real audio (2,000 per dataset). Real samples were never seen during training and serve as hard negatives to simulate out-of-distribution inputs. The ADA classifier achieves a 26.82\% error rate demonstrating effective rejection of anomalous inputs. Among the 5,386 samples classified as \textit{CodecFake}, the ADMR classifier introduces additional errors, with a 35.37\% misclassification rate. These results highlight the impact of early-stage decisions in the LAVA pipeline and confirm the importance of robust attribution at both levels.

\subsection{Generalization Test}
To evaluate the architecture’s capacity for generalization, we tested both classifiers on 20,000 synthetic samples from ASVspoof2019 LA~\cite{wang2020asvspoof2019largescalepublic}, a dataset not used during training, but semantically close to ASVspoof2021, although it includes different spoofing techniques, codec chains, and non-overlapping speaker identities. In the ADA task, 28.82\% of the samples were correctly rejected as unknown. Most of the remaining samples were attributed to ASVspoof2021 (64.18\%), a behavior consistent with the similarity between the two datasets. The rejection mechanism proved effective in isolating anomalous inputs whose confidence scores did not match any known class distribution. In the ADMR task, the model achieved 81.28\% accuracy, demonstrating strong rejection capabilities even under open-set conditions. This result suggests that the latent space learned by the autoencoder, combined with attention-based reweighting, enables robust handling of unseen synthesis techniques that share latent similarities with known codecs.

\subsection{Ablation Studies}
\label{sec:ablation}
To assess the importance of attention mechanisms, we trained variants of both classifiers with the attention modules removed. As shown in Table~\ref{tab:attention_ablation}, performance consistently dropped across all tasks. The impact was especially severe in the ADMR classifier (Table~\ref{tab:attention_ablation}), whose accuracy fell from 96.32\% to 82.56\%. A closer inspection of class-level metrics (Table~\ref{tab:ADMR_ablation}) reveals significant performance degradation, particularly for class F06, whose recall dropped to 42.63\%. This underscores the value of attention for fine-grained attribution, where subtle feature differences must be preserved and leveraged for reliable classification.

% \begin{table}[h]
% \centering
% \caption{Performance comparison with and without attention for ADA and ADMR tasks}
% \label{tab:attention_ablation}
% \resizebox{\columnwidth}{!}{%
% \begin{tabular}{lcccc}
% \toprule
% \textbf{Model} & \textbf{Precision} & \textbf{Recall} & \textbf{F1-Score} & \textbf{Accuracy} \\
% \midrule
% \multicolumn{5}{c}{\textbf{Level 1: ADA}} \\
% \midrule
% With Attention & 0.9625 & 0.9621 & 0.9622 & 0.9621 \\
% Without Attention & 0.9104 & 0.9079 & 0.9082 & 0.9079 \\
% \midrule
% \multicolumn{5}{c}{\textbf{Level 2: ADMR}} \\
% \midrule
% With Attention & 0.9633 & 0.9632 & 0.9631 & 0.9632 \\
% Without Attention & 0.8412 & 0.8256 & 0.8172 & 0.8256 \\
% \bottomrule
% \end{tabular}%
% }
% \end{table}

\begin{table}[t!]
\centering
\caption{Performance comparison with and without attention for ADA and ADMR tasks}
\label{tab:attention_ablation}
    \begin{adjustbox}{max width=.5\textwidth}
\begin{tabular}{cccccc}
\hline
                                                                            \textbf{Level}& \textbf{Model}    & \textbf{Precision} & \textbf{Recall} & \textbf{F1-Score} & \textbf{Accuracy} \\ \hline
\multirow{2}{*}{\resizebox{0.6cm}{!}{\rotatebox[origin=c]{90}{\textbf{\begin{tabular}[c]{@{}c@{}}L1\\ ADA\end{tabular}}}}}  & With Attention    & 0.9625             & 0.9621          & 0.9622            & 0.9621            \\
                                                                            & Without Attention & 0.9104             & 0.9079          & 0.9082            & 0.9079            \\ \hline
\multirow{2}{*}{\resizebox{0.6cm}{!}{\rotatebox[origin=c]{90}{\textbf{\begin{tabular}[c]{@{}c@{}}L2\\ ADMR\end{tabular}}}}} & With Attention    & 0.9633             & 0.9632          & 0.9631            & 0.9632            \\
                                                                            & Without Attention & 0.8412             & 0.8256          & 0.8172            & 0.8256            \\ \hline
\end{tabular}
\end{adjustbox}
\end{table}

\begin{table}[t]
\centering
\caption{ADMR results \textit{without attention}.}
\label{tab:ADMR_ablation}
\resizebox{\columnwidth}{!}{%
\begin{tabular}{lccc}
\toprule
\textbf{Class} & \textbf{Precision} & \textbf{Recall} & \textbf{F1-Score} \\
\midrule
F01 & 0.9605 & 0.9497 & 0.9551 \\
F02 & 0.6372 & 0.8926 & 0.7436 \\
F03 & 0.8687 & 0.8962 & 0.8823 \\
F04 & 0.8389 & 0.8973 & 0.8671 \\
F05 & 0.8771 & 0.8914 & 0.8842 \\
F06 & 0.8650 & 0.4263 & 0.5711 \\
\midrule
Accuracy & \multicolumn{3}{c}{0.8256} \\
Macro Avg & 0.8412 & 0.8256 & 0.8172 \\
Weighted Avg & 0.8412 & 0.8256 & 0.8172 \\
\bottomrule
\end{tabular}
}
\end{table}

%% file: sec/6_discussion.tex
\section{Discussion}
\label{sec:discussion}

The proposed LAVA architecture demonstrates strong attribution performance across both coarse-grained (technology-level) and fine-grained (model-level) tasks. The integration of attention mechanisms proves consistently beneficial, particularly in the more challenging ADMR task, where subtle generator-specific artifacts must be isolated in a shared latent space. The error propagation analysis reveals the importance of accurate predictions at early stages of the pipeline: misclassifications in ADA significantly affect ADMR outcomes, validating the hierarchical structure’s sensitivity to upstream decisions. Our generalization test on ASVspoof2019 LA confirms that the model can extrapolate beyond its training data. In the ADA stage, the model correctly rejects 28.82\% of the samples as unknown and assigns 64.18\% to ASVspoof2021, behavior consistent with the distributional similarity between the two datasets. In the ADMR task, the system achieves 81.28\% accuracy, despite the test set being entirely unseen during training. This highlights the effectiveness of the rejection mechanism in filtering anomalous inputs and the model's capacity to generalize under open-set conditions.

\subsection{Discussion on Prior Works}
While several recent works have addressed tasks related to audio deepfake attribution, none of them perform the same two-level, supervised attribution that LAVA targets. Existing approaches typically focus on either clustering-based identification, attacker recognition, or vocoder tracing, often in closed-set or in-domain conditions. As such, a direct, level-by-level comparison is not possible. Nonetheless, we provide an overview of representative methods that address similar goals from different perspectives. Müller et al.~\cite{müller2022attackerattributionaudiodeepfakes} propose attacker-level neural embeddings trained on ASVspoof2019 and report high accuracy (97.10\%) in a closed-set speaker identification task. Although they explore clustering in an out-of-domain setting by holding out some identities, their method is not designed for generator attribution and lacks any rejection mechanism, key features in forensic attribution tasks. Klein et al. source tracing system~\cite{sourceTracing2024} focuses on reconstructing generation pipelines using acoustic and vocoder model inference. Their goal is to determine the transformation chain behind a spoofed signal, rather than attributing it to a known model. Their reported 84.6\% accuracy refers to closed-set vocoder classification in a highly controlled environment, without open-set evaluation or supervised class-level attribution. Recent methods have also begun exploring open-set scenarios. The TADA framework~\cite{stan2025tada} uses self-supervised embeddings and $k$-NN to cluster audio samples based on generator identity. However, unlike LAVA, TADA does not rely on predefined class labels and does not perform supervised attribution; rather, it attempts to associate each sample with a latent model identity. Its unsupervised clustering nature makes it fundamentally different and not directly comparable. Another recent study, ReTA~\cite{10800741}, introduces a strategy for rejection threshold adaptation in open-set deepfake attribution. While the results on SFR and DFAD datasets are promising, the lack of code, use of uncommon benchmarks, and reliance on static ResNet features without encoder-decoder or attention mechanisms limit comparability with LAVA. In summary, LAVA introduces a unified and structured attribution pipeline that:
\begin{itemize}
    \item supports both in-domain and out-of-domain evaluation;
    \item integrates a confidence-based rejection mechanism for open-set robustness;
    \item enables both technology-level (ADA) and model-level (ADMR) supervised attribution with strong generalization to previously unseen data.
\end{itemize}
This dual-level attribution design is especially important in forensic contexts, where investigators must not only detect synthetic content but also trace its exact origin in terms of the underlying synthesis technology and architecture. These distinctions position LAVA as a reliable and scalable tool for forensic analysis of audio deepfakes in real-world conditions.

%% file: sec/7_conclusions.tex
\section{Conclusions}
\label{sec:conclusion}
In this work, we introduced LAVA, a novel multi-level framework for audio deepfake attribution and model recognition grounded in a shared convolutional autoencoder trained exclusively on synthetic audio. The architecture supports both technology-level attribution (ADA) and fine-grained model recognition (ADMR), and leverages attention mechanisms to enhance performance, while the modular design improves transparency and task specialization.. A key feature of LAVA is its ability to operate in open-set conditions via a confidence-based rejection mechanism that prevents overconfident misclassification of unfamiliar inputs—an essential requirement for forensic deployment. Experimental results on (\textit{CodecFake}, \textit{FakeOrReal}, and \textit{ASVspoof2021}) show that LAVA achieves high attribution performance across tasks, with consistent improvements brought by the attention mechanism and stable behavior under distributional shifts. Our generalization test on the unseen ASVspoof2019 LA dataset confirmed the system's robustness, with 81.28\% accuracy in ADMR and a well-calibrated rejection behavior in ADA. Error propagation analysis emphasized the importance of robust upstream decisions, while ablation studies confirmed the crucial role of attention in capturing model-specific artifacts. Compared to recent approaches, LAVA provides a unique combination of supervised attribution, modularity, and rejection-aware generalization. Unlike prior work that focuses on unsupervised clustering, speaker identity, or vocoder pipeline tracing, LAVA offers direct attribution of synthetic audio to both the generation technology and the underlying model, addressing a pressing need in forensic audio analysis. Its hierarchical and modular design, with clearly defined decision stages, makes it a strong candidate for integration into real-world forensic workflows. Future research will extend the framework with additional attribution levels (e.g., family-level generalization), explore multimodal fusion with visual deepfake detectors, and investigate attribution-aware defenses for online content moderation and forensic auditing.

\section{Acknowledgments}
This study has been partially supported by SERICS (PE00000014), including its vertical project FF4ALL, under the MUR National Recovery and Resilience Plan funded by the European Union – NextGenerationEU.